\documentclass[aps,prx, reprint,longbibliography, nofootinbib,superscriptaddress]{revtex4-1}

\usepackage{subfigure}
\usepackage{graphicx,amsmath,amssymb,color,amsfonts,bm,physics}
\usepackage{mathtools,lipsum}
\usepackage[dvipsnames]{xcolor}
\usepackage[misc]{ifsym}
\usepackage{quantikz}
\usepackage{algorithm}
\usepackage{algpseudocode}
\usepackage[colorlinks=true]{hyperref}  
\hypersetup{
    bookmarks=true,         
    unicode=false,          
    pdftoolbar=true,        
    pdfmenubar=true,        
    pdffitwindow=false,     
    pdfstartview={FitH},    
    pdftitle={},    
    pdfauthor={},     
    pdfsubject={},   
    pdfcreator={},   
    pdfproducer={}, 
    pdfkeywords={} {} {}, 
    pdfnewwindow=true,      
    colorlinks=true,       
    linkcolor=blue, 
    citecolor=blue,        
    filecolor=magenta,      
    urlcolor=blue,           
        breaklinks=true
} 

\newcommand{\be}{\begin{equation}}
\newcommand{\ee}{\end{equation}}
\newcommand{\bea}{\begin{eqnarray}}
\newcommand{\eea}{\end{eqnarray}}

\begin{document}

\title{Quantum Control Geometry as a Metric of Chaos}

\author{Mahmoud Abdelshafy and Joaquin F. Rodriguez-Nieva}
\affiliation{Department of Physics \& Astronomy, Texas A\&M University, College Station, TX 77843}


\begin{abstract}

Quantum-chaotic systems share the same coarse-grained features characteristic of random matrices, such as Wigner–Dyson level-spacing statistics, yet can exhibit strikingly different eigenstate structure and many-body dynamics, including relaxation, transport, scrambling, and entanglement growth. This raises the fundamental question of whether these finer-grained features probe genuinely distinct aspects of quantum chaos or, instead, reflect different manifestations of a common underlying structure---and whether such structure provides a meaningful basis for regarding one many-body quantum system as ``more chaotic'' than another. Here, we address this question by introducing a {\it quantum geometric} framework that quantifies the susceptibility of many-body eigenstates to physically allowed deformations of the Hamiltonian, as constrained by spatial locality, symmetries, and conservation laws. The resulting geometric susceptibility is {\it quantitative}, providing a metric of chaos beyond traditional binary diagnostics; {\it intrinsic}, as it depends on the manifold of allowed Hamiltonian deformations rather than on the way the system is probed; and {\it unifying}, as it directly connects eigenstate structure with dynamics. For canonical models of quantum chaos, we find that minimal geometric susceptibility, maximal eigenstate randomness, broad spectral response of local operators, and the suppression of nonhydrodynamic slow structure coalesce in the same region of Hamiltonian control space. This suggests that these seemingly distinct signatures of chaos are different manifestations of a common underlying geometric structure. Our framework provides a route toward understanding the limits of relaxation and scrambling under physical constraints; in applications, it offers a means to engineer strongly chaotic regimes in programmable quantum systems with restricted controls.

\end{abstract}



\maketitle

\section{Introduction}

Our understanding of quantum chaos in many-body systems has been largely shaped by the dichotomy between generic chaotic systems that obey conventional statistical mechanics and nonchaotic systems that evade thermalization~\cite{DAlessio_review_16,Nandkishore2015,2019RMP_mblreview}. Yet even among systems that ultimately thermalize, not all quantum chaotic systems are chaotic in the same way: their dynamics span a broad continuum, ranging from long-lived prethermal regimes~\cite{2004berges,Abanin_15,2019PRX_Prethermalization,integrabilitybreaking3} to strongly chaotic regimes in which relaxation approaches the fastest timescales allowed by fundamental physical constraints~\cite{PhysRevX.7.031027,2019PRX_lanczos,RevModPhys.94.041002}.

Moving beyond the chaotic-nonchaotic dichotomy requires going beyond {\it qualitative diagnostics} of chaos, such as random-matrix statistics of eigenenergies and eigenstates or matrix elements of local observables~\cite{DAlessio_review_16,Nandkishore2015,2019RMP_mblreview}, toward {\it quantitative metrics} that provide a highly resolved characterization of the dynamical structure of many-body quantum systems. 
This endeavor is important for both fundamental physics and quantum technologies. At a fundamental level, quantitative measures of chaos open the possibility of identifying its extremal limits and understanding how microscopic constraints bound the rates of information scrambling, relaxation, and equilibration. This, in turn, could provide new insight into the timescales governing equilibration in strongly interacting materials and complex molecules~\cite{RevModPhys.94.041002,Leitner04072015} and, more broadly, into the fundamental limits on how rapidly quantum processes can occur in nature~\cite{Deffner_2017}.
In addition, many quantum technologies, ranging from randomized benchmarking~\cite{Choi_23,Mark_23} and quantum tomography~\cite{Aaronson_20,Huang_20} to demonstrations of quantum computational advantage~\cite{Arute2019,Bouland2018,Schuster_25}, benefit from the efficient production of random states. 
This raises the question of how a device with only a few controllable knobs can be configured to maximize or suppress information scrambling, or enhance the generation of entanglement. 

In this spirit, a growing body of work has sought to characterize quantum chaos beyond the universal statistical behavior predicted by random matrix theory. Important advances include understanding how spatial locality and conservation laws influence spectral correlations~\cite{Friedman2019,Bertini_18,Kos_18,Winer2022} and higher-order correlations between eigenstate matrix elements of local observables~\cite{Brenes_21,Hahn2024,Dymarsky_22,PhysRevX.15.011059,Pappalardi_25,2026PRL_ETHHydro}; characterizing the spatial spreading and increasing complexity of simple operators under chaotic dynamics~\cite{Nahum_18,vonKeyserlingk_18,2018PRX_khemani,2019PRX_lanczos}; 
quantifying the effects of perturbations on quantum chaotic eigenstates~\cite{Pandey_20,LeBlond_21,Abdelshafy_25}; identifying subleading corrections to the volume-law entanglement entropy of chaotic eigenstates~\cite{Vidmar_17,Murthy_19,Huang2019,Haque_22,Bianchi_22,Rodriguez-Nieva_24,2025PRL_EEU1}; or developing statistical descriptions that capture higher moments of quantum state ensembles arising from dynamics~\cite{PhysRevX.14.041059,2024PRX_maximumentropyprinciple,ghosh_25}, microcanonical sampling~\cite{Bianchi_22,Rodriguez-Nieva_24,2025PRL_EEU1}, or projective measurements on part of the system~\cite{2022PRL_exactdesigns,Cotler_23,2022Quantum_designs}. 
Combined, these works have revealed a rich fine structure of quantum chaos shaped by the interplay between various microscopic constraints, including spatial locality, symmetries, and conservation laws.

Despite these advances, several fundamental challenges remain. The most basic is the absence of a universally accepted definition of quantum chaos~\cite{DAlessio_review_16,KudlerFlam2020}. While there is broad agreement on its characteristic signatures~\cite{DAlessio_review_16,Nandkishore2015,2019RMP_mblreview}---including random-matrix statistics of eigenenergies, eigenstate thermalization, information scrambling, and dynamical relaxation---the diversity of existing chaos diagnostics~\cite{KudlerFlam2020,Bhattacharyya2022} raises the question of whether they capture different manifestations of a common underlying structure or, instead, whether they characterize fundamentally distinct aspects of quantum chaos. This ambiguity is further compounded by the dependence of many diagnostics on the observable or perturbation used to probe the system~\cite{Pandey_20,LeBlond_21,Abdelshafy_25,Suntajs_23,ABANIN_21}. A Hamiltonian may appear strongly chaotic under one probe while retaining highly structured behavior under another~\cite{Suntajs_23,ABANIN_21,sharipov_26}. 

By the same token, there is no general framework for establishing whether one quantum system should be regarded as ``more chaotic'' than another. 
This gap becomes especially apparent when one attempts to define an extremal notion of chaos---or, colloquially, an ``arrow of chaos'' pointing toward increasingly chaotic behavior. For instance, different communities have used the term {\it maximal chaos} in diametrically opposite ways: on the one hand, to describe systems that thermalize and scramble information at the fastest possible rates~\cite{Maldacena_16,Shenker_14} and, on the other hand, to describe systems that are maximally susceptible to local perturbations, a behavior that is instead prominent near integrable limits~\cite{Pandey_20,Kim_24,Kim_Lim_26}. 

Ultimately, all these challenges reduce to a fundamental question: can we {\it quantify} quantum chaos in a way that is {\it intrinsic} to the Hamiltonian, independent of any particular probe of choice, and {\it unifies} its different manifestations within a common theoretical framework?

In this work, we address this question by developing a {\it quantum geometric} framework that (i) goes beyond binary diagnostics to provide a quantitative measure of quantum chaos, (ii) provides an {\it intrinsic} characterization of the Hamiltonian that does not rely on a particular choice of probe but instead on the geometry of physically allowed Hamiltonian deformations, and (iii) {\it unifies} distinct manifestations of quantum chaos---including eigenstate structure, spectral response, and relaxation dynamics---within a single tensorial quantity. We apply this framework to paradigmatic models of quantum chaos, revealing how the geometry of allowed Hamiltonian deformations connects eigenstate structure to many-body dynamics across different timescales.

\begin{figure}
    \centering
    \includegraphics{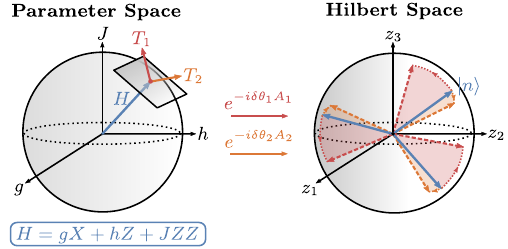}
    \caption{Geometry of Hamiltonian controls and the corresponding eigenstate response under physically allowed deformations. The many-body Hamiltonian $H_{\boldsymbol\theta}=gX+hZ+JZZ$, together with its control operators $X,Z,ZZ$, defines a quantum control manifold (here we use $Z$, $ZZ$, and $X$ as shorthand for the corresponding uniform spin operators). Fixing the overall energy scale, $\frac{1}{\cal D}{\rm Tr}[H^2]=N\varepsilon_*^2$, with $\cal D$ the Hilbert space dimension and $N$ the system size, restricts this manifold to the sphere shown schematically: a radial perturbation simply rescales the Hamiltonian, whereas orthogonal directions $T_\alpha$ change its physical composition while preserving its overall energy scale. 
    Infinitesimal deformations on the orthogonal space rotate the Hamiltonian eigenstates in Hilbert space through the corresponding adiabatic gauge potential (AGP), $|n\rangle\rightarrow e^{-i\epsilon A_\alpha}|n\rangle$. The magnitude and anisotropy of these eigenstate displacements define the geometric susceptibility tensor and quantify how strongly the eigenbasis responds to physically allowed changes of the Hamiltonian.}
    \label{fig:schematics}
\end{figure}

\subsection{Summary of key results}

Specifically, we construct a {\it geometric susceptibility} tensor that quantifies the sensitivity of Hamiltonian eigenstates to all physically allowed deformations of the Hamiltonian. It is geometric in the sense that it depends only on the Hamiltonian and the manifold of allowed deformations, as constrained by locality and symmetries, rather than on a particular choice of probe. As illustrated in Fig.~\ref{fig:schematics}, the Hamiltonian and the allowed deformations define a tangent space that naturally separates the trivial deformation parallel to the Hamiltonian, which merely rescales its energy and leaves the eigenstates unchanged, from orthogonal directions that genuinely deform the eigensystem. The allowed operators are first placed on equal footing through a common Hilbert--Schmidt normalization and orthogonalized with respect to the Hamiltonian. This endows the physical control space with a well-defined geometry, so that susceptibilities along different directions can be meaningfully compared independently of how the Hamiltonian and deformations are parametrized. In this respect, our construction differs from the use of fidelity susceptibility along a preselected perturbation~\cite{Pandey_20,LeBlond_21,Abdelshafy_25}, which would then depend on the details of the probe. Importantly, scalar invariants of the resulting geometric tensor, most simply its trace, quantify the overall susceptibility of the eigensystem, while its eigenvalues and principal deformations resolve the magnitude and anisotropy of the response throughout Hamiltonian control space.

Although the geometric susceptibility is a quasistatic property of the Hamiltonian, it is intimately connected to dynamics: geometric susceptibility and relaxation dynamics are governed by the same matrix elements of physically allowed operators, with the former placing an additional inverse-frequency weight on their spectral response~\cite{2017PhysRep_AGP,2019RMP_shortcutstoadiabaticity}. The geometric tensor thus provides a direct link between eigenstate structure, spectral response, and relaxation dynamics, revealing that these distinct properties of many-body quantum systems are different manifestations of the same underlying geometry. 
In this respect, the geometric tensor provides a direct connection to dynamics, complementing diagnostics based solely on eigenstate structure, such as eigenstate entanglement statistics~\cite{Rodriguez-Nieva_24} or the difficulty of learning a Hamiltonian from an individual eigenstate~\cite{gjonbalaj_choi_26}. In particular, we find that the degree to which an eigenstate is delocalized---and hence randomized---in a local basis is closely tied to how strongly it mixes under a small physical deformation, as well as to how broadly the corresponding operator spectral weight is distributed in frequency and how little weight remains at low frequencies. These qualitatively distinct manifestations of chaos are {\it quantitatively} connected through the same geometric framework.

An important implication of this framework is that it provides a unifying picture for previously-studied maximally chaotic regimes~\cite{Rodriguez-Nieva_24} in terms of {\it minimal} geometric susceptibility,\footnote{Previous works~\cite{Pandey_20,Kim_24,Kim_Lim_26} have associated maximal chaos with {\it large} susceptibility, for instance near integrable limits, where small perturbations induce anomalously strong eigenstate rearrangements and slow dynamics. Here we use a distinct definition consistent with scrambling-based notions in which quantum information is rapidly dispersed among many-body degrees of freedom. These contrasting interpretations underscore the absence of a widely accepted definition of quantum chaos.} thereby connecting complementary eigenstate and dynamical perspectives on quantum chaos. From the eigenstate perspective, minimal susceptibility identifies regimes in which eigenstates have essentially lost all physical structure not protected by the underlying constraints. In this regime, physically allowed Hamiltonian deformations are unable to further remove any remaining eigenstate structure. Equivalently, eigenstates are maximally delocalized in a local basis, such that local perturbations maximally hybridize eigenstates with small microcanonical windows, consistent with our previously introduced notion of maximal chaos based on eigenstate randomness~\cite{Rodriguez-Nieva_24}. 

From the dynamical perspective, the inverse-frequency weighting of the geometric susceptibility makes it particularly sensitive to slow relaxation processes. Its minimization therefore signals the suppression of low-frequency spectral weight typically associated with slow modes. In the maximally chaotic limit, all unprotected slow structure is washed out, leaving only the hydrodynamic modes enforced by conservation laws and locality to govern the geometric susceptibility. This picture connects naturally with an independent proposal using the onset of hydrodynamics to define and bound the local equilibration time~\cite{2025NatPhys_ThermalizationBound}.

Taken together, our results show that minimal eigenstate susceptibility, maximal eigenstate randomness, broad and smooth spectral response, and long-time correlations controlled by hydrodynamics coincide in the same region of Hamiltonian parameter space. These signatures represent only a few among the broader web of quantum-chaos diagnostics~\cite{2020JHEP_webofchaos}, but their convergence suggests that these share a common underlying geometric structure, providing a unified description of several key manifestations of quantum chaos.

\subsection{Outline of Work}

The rest of this manuscript is organized as follows. In Sec.~\ref{sec:formalism}, we introduce our geometric framework and establish the connection between the geometry of many-body eigenstates and the dynamical properties of quantum systems. In Sec.~\ref{Sec:model}, we introduce the quantum-chaotic models considered in this work and their corresponding control parameters. In Sec.~\ref{sec:susceptibility}, we establish the connection between the overall geometric susceptibility and eigenstate randomness. In Sec.~\ref{sec:eigenmodes}, we analyze the principal deformation directions of the geometric susceptibility tensor, connecting the static eigenstate properties discussed in Sec.~\ref{sec:susceptibility} to the dynamical response revealed by their spectral functions. In Sec.~\ref{sec:hydro}, we extend our analysis of the dynamical properties of strongly chaotic systems to operators outside the control manifold, demonstrating the suppression of nonhydrodynamic slow modes and the emergence of a regime in which the remaining slow dynamics are governed by hydrodynamics. Finally, in Sec.~\ref{Sec:Discussion}, we summarize our main results and discuss their broader implications and potential directions for future research.

\section{Formalism}
\label{sec:formalism}

In this section, we begin by briefly reviewing the formalism of adiabatic deformations, referring the reader to Refs.~\cite{2017PhysRep_AGP,2019RMP_shortcutstoadiabaticity} for more detailed treatments. We then introduce the central idea of this work first at an intuitive level, showing how the adiabatic response provides the natural bridge between the structure of many-body eigenstates and many-body quantum dynamics.  
Lastly, we describe our formalism in full detail and define the key geometric quantities used throughout the manuscript.

\subsection{A Brief Review on Adiabatic Deformations}
\label{sec:AGP}

Consider a Hamiltonian $H$ subject to a tunable perturbation $O$ controlled by the parameter $\theta$, $H_\theta = H+\theta O$. The adiabatic gauge potential (AGP) is the generator of the unitary transformation that connects the instantaneous eigenstates of $H_\theta$ as $\theta$ is varied. Its off-diagonal matrix elements are~\cite{2017PhysRep_AGP,2019RMP_shortcutstoadiabaticity}
\be 
\langle m|A_\theta|n\rangle = i\,\langle m|\partial_\theta n\rangle = -i\frac{\langle m|O|n\rangle}{E_m - E_n},
\quad m\neq n ,
\label{eq:AGP}
\ee
where $E_m$ and $E_n$ are the eigenergies $m$ and $n$ and $\ket{m}$ and $\ket{n}$ the corresponding eigenstates. The diagonal matrix elements of $A_{\theta}$ are gauge dependent, as these generate arbitrary phase rotations of individual eigenstates without changing the eigenspaces themselves, and can be chosen arbitrarily; throughout this work, we fix this gauge degree of freedom by setting them to zero. 

The inverse-frequency weighting of $A_\theta$ in Eq.~(\ref{eq:AGP}) makes the AGP an extremely sensitive diagnostic of quantum chaos. Its response is controlled jointly by small energy denominators and by the matrix elements of the perturbation connecting the corresponding eigenstates. In integrable systems, the absence of level repulsion allows many exact or near level crossings, while the matrix elements of generic integrability-breaking perturbations remain highly structured and can retain substantial weight between nearly degenerate states, producing a strongly enhanced AGP response~\cite{Pandey_20,LeBlond_21}. By contrast, in quantum-chaotic systems level repulsion suppresses near-degeneracies, while the perturbation matrix elements are broadly distributed over the eigenbasis rather than concentrated in a sparse set of transitions, thereby producing a strong suppression of the low-frequency weight entering the AGP and a parametrically smaller susceptibility~\cite{Pandey_20,LeBlond_21}.

These qualitatively distinct behaviors can be diagnosed via the fidelity susceptibility $F_\theta$, obtained by squaring and tracing Eq.~(\ref{eq:AGP}), 
\be
F_\theta=\frac{1}{\mathcal D}{\rm Tr}\left[A_\theta^\dagger A_\theta\right] = \int_{-\infty}^{\infty} d\omega \frac{S_O(\omega)}{\omega^2},
\label{eq:fidelitysusceptibility}
\ee
where $S_O(\omega)$ denotes the off-diagonal spectral function of the operator $O$, $S_O(\omega) = \frac{1}{\cal D}\sum_{m\neq n} |\langle m | O |n\rangle|^2 \delta(\omega-E_m+E_n)$, and $\mathcal{D}$ is the Hilbert space dimension. The inverse-frequency weighting in Eq.~(\ref{eq:fidelitysusceptibility}) makes the strong sensitivity of $F_\theta$ to low-frequency spectral weight explicit in the final term. 
Reference~\cite{Pandey_20} first identified $F_\theta$ as an extremely sensitive diagnostic of the integrable-to-chaotic crossover, outperforming conventional probes such as level-spacing statistics because it captures not only spectral correlations but also the structure of the corresponding eigenstates. It has since been studied across a broad range of many-body models~\cite{Dries_21,LeBlond_21,Kim_24,Abdelshafy_25,tokarczyk_26,meinersen_26,mihailescu_26}.

In practice, the AGP must be regularized to suppress contributions from exponentially small level spacings:
\be
\langle m|A_\theta|n\rangle
=
\frac{-i\omega_{mn}}{\omega_{mn}^2+\mu^2}
\langle m|O|n\rangle,
\label{eq:regularizedAGP}
\ee
where $\omega_{mn}=E_m-E_n$ is the difference between eigenenergies. This regularization effectively introduces a finite frequency resolution, which can be interpreted physically as an inverse observation time or experimental resolution. It also connects directly to experimental probes: the geometric susceptibility associated with $A_{\theta}$ can be accessed through Loschmidt-echo protocols involving evolution under slightly different Hamiltonians~\cite{Kolodrubetz_13}, or reconstructed from frequency-resolved correlation spectra of the corresponding physical operators~\cite{PhysRevB.97.201117}. 
In this sense, the regularization scale $\mu$ sets the frequency window over which the AGP is experimentally resolved. 

The regularization, Eq.~(\ref{eq:regularizedAGP}), preserves the characteristic odd-in-frequency structure of the AGP. In the limit $\omega_{mn}^2\gg\mu^2$, Eq.~(\ref{eq:regularizedAGP}) reduces to Eq.~(\ref{eq:AGP}), whereas matrix elements within the frequency window $|\omega_{mn}|\lesssim\mu$ are suppressed. The diagonal matrix elements vanish automatically, consistent with our gauge choice. Although $A_\theta$ depends explicitly on the regularization scale $\mu$, this dependence can be scaled out of the susceptibility, leaving its overall magnitude insensitive to the particular choice of $\mu$, as discussed in Sec.~\ref{subsec:formalism}.

\subsection{Connection between eigenstate mixing, eigenstate randomness, and dynamics across scales}

Equation~(\ref{eq:AGP}) provides the starting point for our analysis and admits a simple physical interpretation to which we will return throughout the manuscript. In the spirit of {\it quantifying} chaos, it is instructive to ask what it means for the susceptibility to a physical perturbation to become progressively smaller. The first equality in Eq.~(\ref{eq:AGP}) offers an eigenstate perspective: it quantifies how strongly the eigenbasis rotates under an infinitesimal physical deformation of the Hamiltonian, or equivalently, how a given eigenstate admixes other eigenstates under such a perturbation. 
In regimes of small, potentially minimal, eigenstate susceptibility, physically allowed perturbations couple each eigenstate to a broad set of eigenstates over the accessible energy range. This broadly distributed mixing is consistent with eigenstates that are already nearly maximally random within the constraints of the system, leaving little residual structure that can be further erased by small Hamiltonian deformations. 
This suggests that maximal randomness of individual eigenstates and minimal sensitivity to physical deformations are complementary manifestations of the same underlying structure. 
We are thus led to the central expectation that the regime of {\it minimal} eigenstate susceptibility to physically allowed perturbations coincides with regimes of {\it maximal} eigenstate randomness, as quantified, for instance, by comparing microcanonical distributions of eigenstate entanglement with constrained random matrix predictions~\cite{Rodriguez-Nieva_24}.

The second equality in Eq.~(\ref{eq:AGP}) establishes a direct link between adiabatic deformations and dynamics. The AGP and dynamical correlation functions are governed by the same matrix elements, but the former weights their contributions by inverse powers of frequency, making it particularly sensitive to slow dynamical processes. From this perspective, a decreasing susceptibility is consistent with the suppression of low-frequency spectral weight and its redistribution over a broader frequency range. 
Thus, this should be reflected in broadly distributed, small matrix elements of local operators, giving rise to {\it smooth} and {\it broad} spectral functions, while level repulsion further suppresses contributions from anomalously small energy denominators. 
However, spatial locality and conservation laws impose unavoidable constraints on the relaxation of local observables, giving rise to hydrodynamic modes that ultimately govern the spectral weight at low frequencies~\cite{PhysRevX.15.011059,2025NatPhys_ThermalizationBound}. In the regime of minimal susceptibility, we therefore expect hydrodynamics to provide the only remaining source of slow relaxation, while all dynamical structure not protected by fundamental physical constraints is effectively washed out.

These observations motivate the central conjecture of this work, connecting regimes of minimal eigenstate susceptibility to regimes of maximal eigenstate randomness, broad short-time spectral response, and hydrodynamics-limited late-time dynamics:
\be
\boxed{
\begin{gathered}
\text{Minimal geometric susceptibility}\\
\Updownarrow\\
\text{Maximal eigenstate randomness}\\
\Updownarrow\\
\text{Broad spectral response \&} \\ \text{hydrodynamics-limited correlations}
\end{gathered}
}
\label{eq:hypothesis}
\ee

To make these connections precise and quantitative, we now proceed to transform the eigenstate susceptibility ideas discussed in Sec.~\ref{sec:AGP} from a {\it diagnostic} of quantum chaos into a {\it metric} of chaos. 

\subsection{Geometric Tensor of Control Parameters}
\label{subsec:formalism}

We now generalize the physical picture of the previous subsections to a high-dimensional setting that connects the geometry of physically allowed Hamiltonian terms to the eigenstate and dynamical properties of a given Hamiltonian. To this end, we parameterize the Hamiltonian as
\be
H_{\bm\theta}=\sum_{\alpha=1}^{M}\theta_\alpha L_\alpha,
\label{eq:Ham}
\ee
where $L_\alpha$ denotes the $M$ operators spanning the accessible Hamiltonian manifold. These may be understood as the local operators allowed by the symmetries of the system or, in programmable quantum platforms, as the set of experimentally controllable knobs. Throughout this work, we consider local spin-$\frac{1}{2}$ Hamiltonians, or equivalently qubit systems, and choose the operators $L_\alpha$ to form an orthonormal basis according to 
$\frac{1}{\cal D}{\rm Tr}\left[L_\alpha^\dagger L_\beta\right]=N\delta_{\alpha\beta}$,
where $\mathcal D=2^N$ is the Hilbert-space dimension and $N$ is the number of spins.

For example, for the uniform mixed-field Ising model (MFIM) that will be considered below, the operator basis is $L_\alpha\in\{X,Z,ZZ\}$, so that $M=3$ (Fig.~\ref{fig:schematics}). These three operators control, respectively, the transverse field $g$, the longitudinal field $h$, and the exchange coupling $J$ appearing in the Hamiltonian. One could also assume local controls on each individual spin, which then increases the parameter space to $M = 3N$ terms. Here we will focus on the former scenario for simplicity, but the formalism applies equally well to the latter case. 

The Hamiltonian defines a microscopic energy scale $\varepsilon_*$ through $\frac{1}{\mathcal D}\operatorname{Tr}[H^2]=N\varepsilon_*^2$. 
With the definition of $H_{\bm\theta}$ introduced in Eq.~(\ref{eq:Ham}), this scale is the magnitude of the Hamiltonian vector in parameter space, as illustrated schematically in Fig.~\ref{fig:schematics},
\be
\varepsilon_*=\sqrt{\sum_{\alpha=1}^{M}\theta_\alpha^2}.
\ee
The quantity $\varepsilon_*$ sets the natural microscopic energy scale of the system. Rescaling $H_{\bm\theta}$ at a fixed direction in parameter space changes only the overall energy scale, leaving the eigenstates and the level-spacing statistics expressed in units of $\varepsilon_*$ unchanged. Throughout this work, we therefore express energies and frequencies in units of $\varepsilon_*$, and times in units of $\varepsilon_*^{-1}$, effectively setting $\varepsilon_*=1$ and $\frac{1}{\mathcal D}\operatorname{Tr}[H^2]=N$. 

The Hamiltonian $H_{\bm\theta}$, together with the manifold of controllable Hamiltonian operators, defines the {\it geometric susceptibility} tensor 
\be
\left[ G_{\bm\theta} \right]_{\alpha\beta} = \frac{1}{\mathcal D}{\rm Tr}\left[A_\alpha^\dagger A_\beta\right]
= \int_{-\infty}^{\infty} d\omega 
\frac{\omega^2 S_{\alpha\beta}(\omega)}{(\omega^2+\mu^2)^2} ,
\label{eq:geometrictensor}
\ee
where $A_\alpha$ is the (regularized) AGP associated with the operator $L_\alpha$, thus generalizing Eq.~(\ref{eq:fidelitysusceptibility}) to a high dimensional operator manifold. In Eq.~(\ref{eq:geometrictensor}), we have implicitly specialized to infinite temperature, thereby weighting the matrix elements of the AGP from all eigenstates equally. This is the setting considered throughout this work, although finite-temperature or microcanonical versions of the geometric tensor can be defined straightforwardly by incorporating the corresponding statistical weight of each eigenvector. The second equality in Eq.~(\ref{eq:geometrictensor}) follows from Eq.~(\ref{eq:regularizedAGP}), with $S_{\alpha\beta}(\omega) = \frac{1}{\cal D}\sum_{m\neq n} \langle m|L_\alpha|n\rangle
\langle n|L_\beta|m\rangle \delta(\omega-E_m+E_n)$ 
denoting the off-diagonal spectral function at infinite temperature.

The geometric tensor defined in Eq.~(\ref{eq:geometrictensor}) provides the desired extension from an operator-dependent diagnostic of chaos to a geometric, operator-agnostic metric of chaos. First, it characterizes the response of the entire manifold of physically allowed Hamiltonian deformations rather than that associated with a particular perturbing operator. Its representation depends on the choice of operator basis, but its geometric content---including its eigenvalues, principal directions, and basis-invariant quantities such as \({\rm Tr}[G_{\bm\theta}]\)---is unchanged under transformations between orthonormalized bases. Second, for a general perturbation $O=\sum_\alpha \lambda_\alpha L_\alpha$, the corresponding susceptibility is simply the quadratic form  $\chi_O=\sum_{\alpha\beta} \lambda_\alpha \left[G_{\bm\theta}\right]_{\alpha\beta}\lambda_\beta$, thus $G_{\bm\theta}$ encodes the susceptibility along every direction in the control manifold while ${\rm Tr}[G_{\bm\theta}]$ averages over the orthonormal set of deformations. 
Third, by expressing frequencies in units of $\varepsilon_*$, we eliminate the trivial radial direction associated with changing the overall microscopic energy scale, so that the resulting metric only probes structural changes of the Hamiltonian rather than trivial rescalings of its energy scale. 
All these desirable features follow from placing the allowed operators on equal footing through a common Hilbert--Schmidt normalization and orthogonalizing them with respect to the Hamiltonian, so that susceptibilities along different directions can be meaningfully compared independently of how the Hamiltonian and its deformations are parametrized. 

The geometric tensor defines a spectrum ${\mathfrak g}_{\bm\theta,\alpha}$ and a set of eigendirections $ T_\alpha= \sum_{\beta=1}^M [v_{\bm\theta,\alpha}]_\beta L_\beta$ determined by the available Hamiltonian controls (Fig.~\ref{fig:schematics}). One of these directions has necessarily zero eigenvalue and is associated with the Hamiltonian $H_{\bm\theta}$ itself, with corresponding eigenvector $[v_{\bm\theta,0}]_\beta=\theta_\beta/\varepsilon_*$ ($\beta = 1,2,\ldots,M$).
The other directions have non-zero contributions to the geometric tensor. 
We use the total geometric susceptibility of the Hamiltonian $H_{\bm\theta}$, 
\be
{\mathfrak g}_{\bm\theta}={\rm Tr}[G_{\bm\theta}],
\ee
which we will denote more compactly as the {\it geometric susceptibility}, 
as a quantitative measure of chaos, quantifying the total susceptibility of the Hamiltonian to all physically allowed deformations. 

As noted above, the magnitude of $\mathfrak g_{\bm\theta}$ depends on the regularization scale $\mu$. However, $\mu$ primarily sets the overall scale of the susceptibility without altering its relative variation across the Hamiltonian control space.  
To see this, consider a quantum chaotic system in which the spectral functions of one particular deformation can be approximated as a broad, smooth Lorentzian function, $S_\alpha(\omega)\simeq \frac{N}{\pi}\frac{\Gamma_{\bm\theta,\alpha}}{\omega^2+\Gamma_{\bm\theta,\alpha}^2}$ (for a finite-bandwidth many-body spin Hamiltonian, the spectral function necessarily deviates from a Lorentzian at frequencies approaching the microscopic bandwidth~\cite{2019PRX_lanczos,PhysRevResearch.2.043234}, but these high-frequency details do not affect the leading behavior of $\mathfrak g_{\bm\theta}$). Substituting this form into Eq.~(\ref{eq:geometrictensor}) gives 
\be
{\mathfrak g}_{\bm\theta,\alpha} = \frac{N\Gamma_{\bm\theta,\alpha}}{2\mu(\Gamma_{\bm\theta,\alpha}+\mu)^2} 
\approx\frac{N}{2\mu\Gamma_{\bm\theta,\alpha}},
\label{eq:scaling}
\ee
where in the last step we used $\mu \ll \Gamma_{\bm\theta,\alpha}\approx \mathcal{O}(\varepsilon_*)$. As such, in the quantum chaotic regime, one can make the total geometric susceptibility independent of $\mu$ and $N$ by rescaling and taking the limit
\be
\tilde{\mathfrak g}_{\bm\theta,\alpha} = \lim_{\mu\rightarrow 0} \lim_{N\rightarrow\infty}\frac{2\mu{\mathfrak g}_{\bm\theta,\alpha}}{N}.
\ee
For the Lorentzian example above, $\tilde{\mathfrak g}_{\bm\theta,\alpha}  = \Gamma_{\bm\theta,\alpha}^{-1}$, the rescaled susceptibility measures the normalized zero-frequency spectral density, which in the Lorentzian approximation is equivalent to the inverse spectral width of mode $\alpha$. 

We note that the scaling in Eq.~(\ref{eq:scaling}) breaks down near integrable limits, where spectral functions are no longer broad and smooth. In particular, in finite-sized numerics, the region where this breakdown occurs shrinks with increasing system size~\cite{Pandey_20,LeBlond_21,Abdelshafy_25} (see also Appendix~\ref{app:finitesizescaling}), so that the scaling in Eq.~(\ref{eq:scaling}) holds everywhere except on the integrable lines in the thermodynamic limit.

Having defined the formalism, we note that the geometric tensor identifies a special point $\bm\theta_*$ in parameter space at which the total geometric susceptibility ${\mathfrak g}_{\bm\theta}$ is minimized. In particular, whereas the geometric susceptibility can become arbitrarily large for increasing system size near integrable limits, it cannot decrease indefinitely and instead reaches a minimum at
\be
\bm\theta_* = \underset{\bm\theta\in{\rm chaotic}}{\rm arg\,min}\,\,[\tilde{\mathfrak g}_{\bm\theta}].
\label{eq:minimization}
\ee
This point in parameter space admits a natural physical interpretation: physically allowed perturbations are unable to remove any further structure from the eigenstates, indicating that they are maximally randomized within the constraints imposed by locality and symmetries such that eigenstates couple broadly across the accessible eigenstate manifold. 
Equivalently, $\bm\theta_*$ identifies the point at which the characteristic spectral broadening $\Gamma_{\bm\theta}$, averaged over deformation directions orthogonal to the Hamiltonian, is maximal.  
This is consistent with generic relaxation on microscopic timescales, with the only slower contributions arising from overlap with hydrodynamic modes. 

Through the next few sections, we explicitly explore the connections between minimal geometric susceptibility and the relations summarized in Eq.~(\ref{eq:hypothesis}). In Sec.~\ref{sec:susceptibility}, we begin by examining the correspondence between $\bm\theta_*$ and the points of maximal eigenstate randomness, as defined in Ref.~\cite{Rodriguez-Nieva_24}, supporting the interpretation that eigenstates at $\bm\theta_*$ are already maximally randomized within the symmetry- and locality-constrained Hilbert space. 
In Sec.~\ref{sec:eigenmodes}, we show that the subspace orthogonal to the Hamiltonian at $\bm\theta_*$ is characterized by an approximately isotropic response, with broadly distributed spectral functions and characteristic widths $\Gamma_{\bm\theta}\sim \mathcal{O}(\varepsilon_*)$. This shows that generic physical perturbations relax on the microscopic timescale set by $\varepsilon_*^{-1}$, largely independently of their direction in Hamiltonian space. In Sec.~\ref{sec:hydro}, we strengthen this picture by showing extensive evidence that hydrodynamic modes provide the only source of parametrically slow relaxation at the susceptibility minimum. This is shown by extending the analysis beyond the operators contained in the control basis and considering the set of inhomogeneous perturbations.

\section{Model \& Control Parameters}
\label{Sec:model}

We consider the one-dimensional mixed field Ising model (MFIM), 
\be
H = \sum_i J Z_iZ_{i+1} + gX_i + h Z_i,
\label{eq:MFIM}
\ee
which has been widely studied as a canonical model of quantum chaos~\cite{2013PRL_Huseentanglementgrowth,2011PRL_banulscirachastings}. Here ${X}_i, {Y}_i, {Z}_i$ denote Pauli matrices at lattice site $i$, $J$ is the exchange coupling, $g$ is the transverse field, and $h$ is the longitudinal field. As shown in Fig.~\ref{fig:schematics}, this defines a three dimensional control space, with one direction determined by the Hamiltonian plus a two-dimensional orthogonal manifold.

In this work, we will consider both open and periodic boundary conditions. In Secs.~\ref{sec:susceptibility} and \ref{sec:eigenmodes}, we use open boundary conditions to match the setup of previous work~\cite{Rodriguez-Nieva_24} and enable a direct one-to-one comparison. Open boundaries explicitly break translation symmetry, while additional boundary fields, $h_1=h_b \hat Z_1$ and $h_L=-h_b \hat Z_N$, break inversion symmetry. Following Ref.~\cite{Rodriguez-Nieva_24}, we take $h_b=0.25$. In Sec.~\ref{sec:hydro}, by contrast, we use periodic boundary conditions when studying inhomogeneous operators, since they allow the operator basis to be organized in momentum space. The choice of boundary conditions does not lead to any qualitative change in the behavior of $G_{\bm\theta}$, although we do observe small quantitative shifts in the location of $\bm\theta_*$. Such shifts are expected, since changing the boundary conditions slightly modifies the finite-size Hamiltonian itself, yet they remain within the uncertainty scales reported below.

As a side technical comment, when computing the geometric tensor and spectral functions, it is necessary to account for the open boundaries in the definition and normalization of the control operators $L_\alpha$ ($\alpha = X,Z,ZZ$). In particular, the missing boundary bond modifies the normalization of the nearest-neighbor $ZZ$ operator, while the boundary fields are incorporated into the definition of the $Z$ operator. The corresponding energy scale is 
$\varepsilon_*=\sqrt{J^2+g^2+h^2+(2h_b^2-J^2)/N}$.
These boundary corrections vanish in the thermodynamic limit, but retaining them keeps the finite-size formalism internally consistent. The explicit orthonormalized operators are given in  Appendix~\ref{app:OBC}.

In our numerical results, we restrict the analysis to a narrow microcanonical window in the middle of the spectrum in order to minimize spectral-edge effects arising in finite-sized systems. Specifically, we use 10\% of the eigenstates centered in the middle of the spectrum, both to compute the geometric tensor and spectral functions. 

For the regularization of the geometric tensor, we choose $\mu$ to be of the order of the typical many-body level spacing, $\Delta\varepsilon = \varepsilon_* \sqrt{2\pi N}/2^N$ for open boundary conditions and $\Delta\varepsilon = \varepsilon_*\sqrt{2\pi N^{3/2}}/2^N$ within each momentum sector for periodic boundary conditions. 
This makes the regularization scale to be much smaller than the intrinsic microscopic energy scale, $\mu \ll \varepsilon_*$, but also to any other slow mode. Since energy is the only conserved quantity in the chaotic regime of the MFIM, $\mu$ satisfies $\mu\sim\Delta\varepsilon\ll D/N^2\ll\varepsilon_*$, where $D$ is the energy diffusion constant. It also naturally leads to the limit $\mu\rightarrow 0$ when $N\rightarrow\infty$, but keeping $\mu$ larger than the level spacing statistics. 
We find that our numerical results are well converged in the quantum chaotic regime for $\mu=5\,\Delta\varepsilon$ for both open and periodic boundary conditions, as shown in the Appendix~\ref{app:regularization}, and use this value throughout. For consistency, we use the same broadening when computing spectral functions $S_{\alpha\beta}(\omega) = \sum_{m\neq n} \langle m|L_\alpha|n\rangle
\langle n|L_\beta|m\rangle \delta(\omega-E_m+E_n)$, replacing the $\delta$-functions by Lorentzians of width $\mu$.

\section{Geometric Susceptibility and eigenstate randomness}
\label{sec:susceptibility}

\begin{figure}
    \centering
    \includegraphics{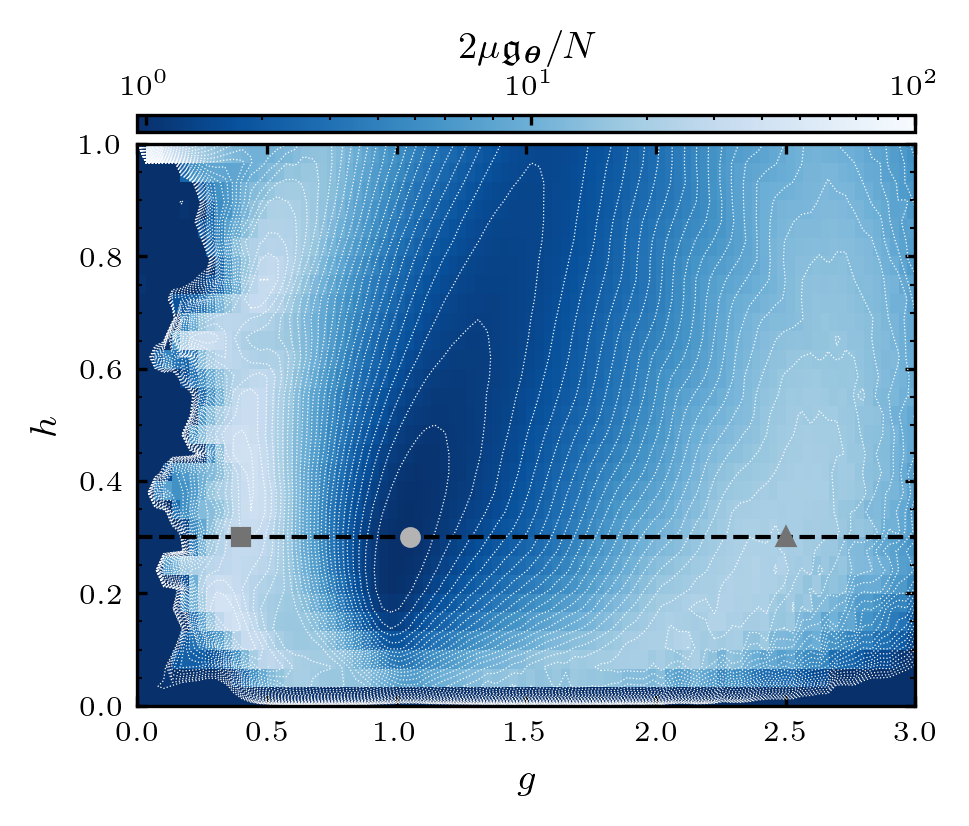}
    \vspace{-4mm}
    \caption{Total geometric susceptibility ${\mathfrak g}_{\bm\theta}={\rm Tr}{[G_{\bm\theta}]}$ across the $(g,h)$ plane in parameter space of the mixed-field Ising model in Eq.~(\ref{eq:MFIM}). We rescale the total susceptibility as $\tilde{\mathfrak g}_{\bm\theta} = 2\mu{\mathfrak g}_{\bm\theta}/N$, which thus becomes independent of the regularization scale and system size in the asymptotic limit (away from integrable limits) and, according to Eq.~(\ref{eq:scaling}), corresponds to the inverse characteristic spectral broadening $\Gamma_{\bm\theta}^{-1}$. 
    The horizontal black dashed line denotes the $h=0.3$ line cut used for Figs.~\ref{fig:G_distribution} and \ref{fig:Geig}. The marked circle denotes the maximally chaotic point $\bm\theta_*$ identified by Eq.~(\ref{eq:minimization}), which is obtained by minimizing the total geometric susceptibility, and agrees remarkably well with the independently obtained point in Ref.~\cite{Rodriguez-Nieva_24} through subsystem-entanglement diagnostics. The gray square and triangle denote the parameters used for Fig.~\ref{fig:G_distribution}. The results are obtained for $J=1$, and $N=14$, and by averaging over the central $10\%$ eigenstates.}
    \label{fig:AGPColormap}
\end{figure}

We begin our analysis by considering the total geometric susceptibility ${\mathfrak g}_{\bm\theta} = {\rm Tr}[G_{\bm\theta}]$ of the MFIM. To enable a direct one-to-one comparison with the results of Ref.~\cite{Rodriguez-Nieva_24}, we project the geometric susceptibility at each point of the unit sphere (Fig.~\ref{fig:schematics}) onto the $(g,h)$ plane while fixing $J=1$, as illustrated in Fig.~\ref{fig:schematics}, and use open boundary conditions. To obtain a quantity that is independent of both system size and the regularization scale in the chaotic regime, we plot the rescaled susceptibility $\tilde{\mathfrak g}_{\bf\theta}=\frac{2\mu\,{\mathfrak g}_{\bm\theta}}{N}$, which, according to Eq.~(\ref{eq:scaling}), corresponds to the characteristic inverse broadening, ${\Gamma}_{\bm\theta}^{-1}$. The independence of $\tilde{\mathfrak g}_{\bm\theta}$ from $\mu$ and $N$ in the quantum chaotic regime is shown in Appendices~\ref{app:regularization} and~\ref{app:finitesizescaling}, respectively.

The resulting geometric susceptibility landscape is shown in Fig.~\ref{fig:AGPColormap}. We find a pronounced minimum developing around $(g_*,h_*)=(1.05\pm0.05,0.3\pm 0.1)$, where the corresponding broadening reaches $\Gamma_{\bm\theta}\approx 1.0\,\varepsilon_*$. The uncertainty is obtained by studying changes in the position of the minimum as a function of system size and regularization scale $\mu$. 
We find that this value agrees remarkably well with the point identified in Ref.~\cite{Rodriguez-Nieva_24} as maximally chaotic through the lens of eigenstate randomness, $(g,h) = (1.1\pm0.05,0.35\pm0.05)$, as quantified by the statistics of entanglement across eigenstates. 
This agreement is nontrivial, as entanglement statistics characterize the structure of individual eigenstates, whereas geometric susceptibility probes energy-resolved matrix elements between different eigenstates under physically allowed deformations. Their coincident extrema suggest that the suppression of residual structure within individual eigenstates is accompanied by a suppression of low-frequency channels through which these deformations mix eigenstates. This provides numerical support for the correspondence proposed in Eq.~(\ref{eq:hypothesis}), with minimal susceptibility coinciding with maximal eigenstate randomness consistent with the constraints imposed by spatial locality and energy conservation.

As we depart from ${\bm\theta}_*$, the geometric susceptibility becomes strongly enhanced near the integrable limits: along the vertical axis, $g=0$, where the Hamiltonian approaches the classical Ising limit, and along the horizontal axis, $h=0$, where the model reduces to the integrable transverse-field Ising chain~\cite{PFEUTY197079}. In these regimes, the response is typically dominated by one or a few deformation directions that are especially efficient at breaking the underlying integrable structure.  
We will return to this point in the next section, where we resolve the individual eigenvalues and eigendirections of $G_{\alpha\beta}$. 

To refine the connection between the geometric susceptibility, which captures the eigenstate-averaged response, and the susceptibility of individual eigenstates, we resolve the geometric susceptibility eigenstate by eigenstate. In the picture developed above, a deformation of a strongly chaotic Hamiltonian mixes each eigenstate broadly across the surrounding spectrum, such that the matrix elements of $A_\alpha$ are distributed over many states in a manner that is largely insensitive to the particular eigenstate considered. The eigenstate-resolved susceptibilities should therefore self-average and become narrowly distributed around ${\mathfrak g}_{\bm\theta}$. This is consistent with the eigenstate thermalization hypothesis~\cite{Srednicki_94,Rigol_08} picture in which neighboring eigenstates become statistically similar and eigenstate-to-eigenstate fluctuations are suppressed. By contrast, near integrable limits, eigenstates hybridize preferentially with a sparse set of nearly resonant states, leading to large fluctuations in susceptibility from one eigenstate—and from one perturbation direction—to another. 

\begin{figure}
    \centering
    \includegraphics{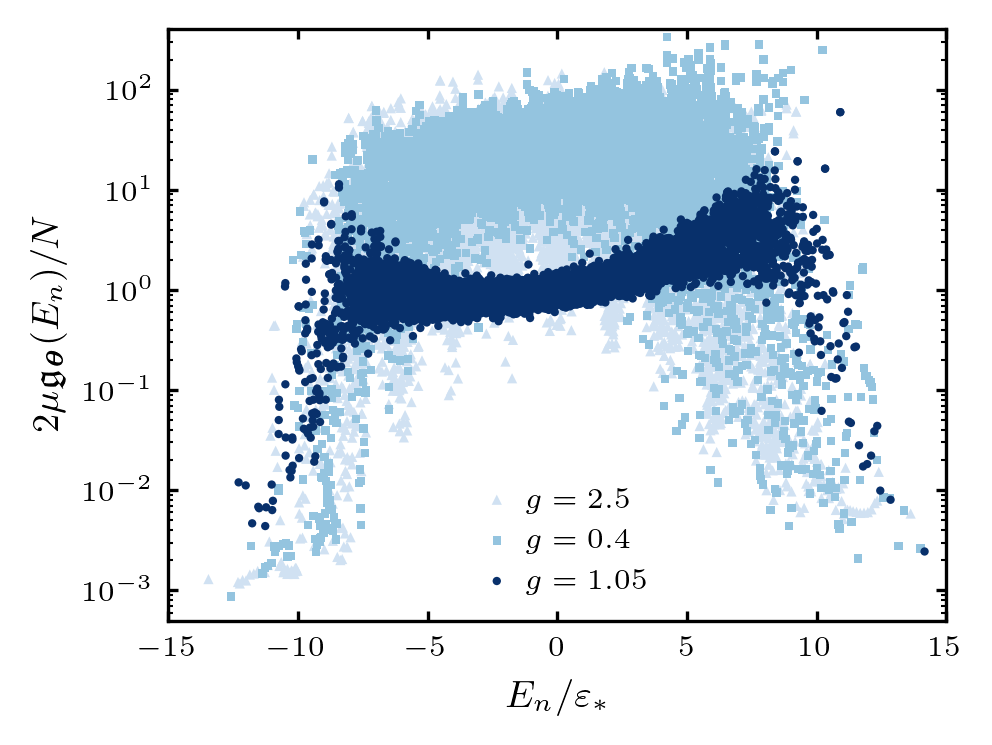}
    \vspace{-4mm}
    \caption{Geometric susceptibility for individual eigenstates, ${\mathfrak g}_{\bm\theta}(E_n)$, rescaled with the regularization and system size. Data points are shown for $g=0.4,1.05,2.5$ along the $h=0.3$ line cut, see Fig.~\ref{fig:AGPColormap}, representing the maximally chaotic point at $g_*=1.05$ and the two regions of enhanced susceptibility around it. The results are for $N=14$ and $J=1$.}
    \label{fig:G_distribution}
\end{figure}

This expectation is borne out in Fig.~\ref{fig:G_distribution}, which shows the distribution of the state-resolved susceptibility $\mathfrak g_{\bm\theta}(E_n)=\sum_{\alpha=1}^{M}\langle n|A_\alpha^\dagger A_\alpha|n\rangle$. At $\bm\theta_*$, the susceptibilities of individual eigenstates are narrowly distributed around a common value, supporting the use of the eigenstate-averaged geometric susceptibility as a representative measure of the response. As the Hamiltonian is tuned away from $\bm\theta_*$, the distribution broadens and increasingly strong state-to-state fluctuations emerge. This growing heterogeneity is consistent with the reappearance of additional eigenstate structure as the system approaches integrability, as discussed above.

\section{Geometric Tensor and the orthogonal operator manifold}
\label{sec:eigenmodes}

We now turn to resolving $G_{\alpha\beta}$ in Eq.~(\ref{eq:geometrictensor}) into its principal deformations. As illustrated in Fig.~\ref{fig:schematics}, for the three-parameter MFIM, the $3\times3$ tensor has one exactly vanishing eigenvalue, whose eigenvector is parallel to the Hamiltonian itself, corresponding to a trivial rescaling of the overall energy scale. The remaining two deformations are orthogonal to the Hamiltonian and carry finite geometric susceptibility. We first discuss the spectral properties of the geometric tensor, followed by its eigendirections and the corresponding spectral functions.

\subsection{Spectrum of the Geometric Tensor}

\begin{figure}
    \centering
    \includegraphics{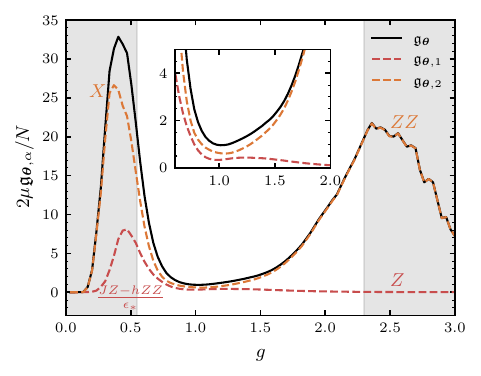}
    \caption{Principal susceptibilities along the $h=0.3$ line cut through parameter space indicated in Fig.~\ref{fig:AGPColormap}. The solid black line shows the corresponding line cut of the normalized total susceptibility, $\tilde{\mathfrak g}_{\bm\theta} = 2\mu\mathcal {\mathfrak g}_{\boldsymbol\theta}/N$, while dashed colored lines denote the non-zero normalized eigenvalues $2\mu {\mathfrak g}_{\bm\theta,\alpha}/N$ for the orthogonal eigendeformations $T_1$ and $T_2$ (the eigenvalue associated to the Hamiltonian vanishes identically). The remaining principal susceptibilities exhibit distinct parameter dependence, but become collectively small near the maximally chaotic regime, producing the pronounced minimum of the total susceptibility. (inset) Zoomed-in version of the main plot around the geometric susceptibility minimum.} 
    \label{fig:Geig}
\end{figure}

In Fig.~\ref{fig:Geig}, we plot the two nonzero eigenvalues of $G_{\bm\theta}$ together with the total geometric susceptibility along the horizontal line cut at $h_*=0.3$, indicated in Fig.~\ref{fig:AGPColormap}, showing the minimum susceptibility at $g_*=1.05$. We divide the plot into two regions: an unshaded region deep in the quantum-chaotic regime, and a shaded crossover region near the integrable limits, where finite-size eigenstates remain influenced by the nearby integrable structure. 
These shaded regions shrink with increasing system size (see Appendix~\ref{app:finitesizescaling}). 

Two features are worth emphasizing. Near the susceptibility minimum, both rescaled susceptibilities are of order $\varepsilon_*^{-1}$, corresponding, within the spectral description above, to the spectral broadening $\Gamma_{\bm\theta,\alpha}\sim\varepsilon_*$. In addition, near $g_*\approx1.05$, the geometric tensor becomes approximately isotropic, $G_{\alpha\beta}\simeq g_0\delta_{\alpha\beta}$, in a Hilbert--Schmidt orthonormal basis of allowed deformations orthogonal to the Hamiltonian. The coincidence of minimal susceptibility and approximate isotropy is surprising, as minimizing the trace of the geometric tensor does not require its nonzero eigenvalues to be approximately equal, and isotropy alone does not imply a small total susceptibility. Thus, for minimal geometric susceptibility, the response to normalized physical deformations is both small and nearly independent of direction. This is consistent with the isotropy of the ensemble-averaged geometric tensor for equivalent perturbations in random-matrix models~\cite{PhysRevLett.126.200604}. 

As one departs $g_*$ and approaches the integrable regimes, the pronounced susceptibility peaks observed in these regions are finite-size precursors of the integrable-to-chaotic crossover and constitute the same characteristic fingerprints emphasized in previous works~\cite{Pandey_20,Dries_21,LeBlond_21,Kim_24,Abdelshafy_25,tokarczyk_26,meinersen_26,mihailescu_26}: with increasing system size, the peak heights grow rapidly while their locations shift closer to the integrable lines. In Appendix~\ref{app:finitesizescaling}, we show the corresponding finite-size analysis, demonstrating that the susceptibility in the unshaded central region has essentially converged for the system sizes considered, whereas the high-susceptibility crossover regions become progressively narrower and more susceptible to perturbations. 

Interestingly, the anisotropy between the two principal susceptibilities grows as the system approaches the integrable limits~\cite{Kim_2026}. Such directional dependence is familiar from adiabatic state-preparation protocols~\cite{Albash_18}, which often begin from classical Hamiltonians with easy-to-prepare eigenstates. At finite driving rates, strongly susceptible deformation directions enhance transitions between instantaneous eigenstates. Efficient protocols thus exploit this anisotropy by favoring weakly susceptible directions, thereby suppressing nonadiabatic transitions and helping the evolving state track the target instantaneous eigenstate~\cite{PhysRevResearch.3.013102}.

\subsection{Eigendeformations of the Geometric Tensor}
\label{sec:eigendirections}

The principal deformations can be obtained analytically near integrable limits, providing limiting cases that help interpret their behavior across the intermediate quantum chaotic regime. Near the Ising limit $g \rightarrow 0$, the Hamiltonian becomes $H \simeq {hZ+JZZ}$, where we use $Z$, $ZZ$, and $X$ as shorthand for the corresponding uniform spin operators, with the appropriate boundary terms understood. The eigenstates of this classical Hamiltonian are product states in the $Z$ basis and can be labeled by their total magnetization and number of domain walls. The two orthonormal eigendirections are $T_1\simeq\frac{JZ-hZZ}{\varepsilon_*}$ and $T_2\simeq X$, with $\varepsilon_*\approx\sqrt{h^2+J^2}$. 
The first operator is itself diagonal in the classical eigenbasis: it shifts and splits energy levels but does not hybridize the eigenstates or alter their underlying product-state structure. Its associated susceptibility is therefore small, as seen in the left shaded region of Fig.~\ref{fig:Geig}. By contrast, $T_2$ flips individual spins and couples states with different magnetization and domain-wall content, thereby introducing genuinely quantum hybridization between the classical sectors. Near the integrable limit, where numerous exact and near degeneracies are present, $T_2$ produces a strongly enhanced susceptibility. The large anisotropy of the geometric tensor therefore directly reflects the very different ability of the two deformation directions to destroy the underlying integrable structure.

A similar argument applies in the paramagnetic limit, $g \rightarrow\infty$, where $H \simeq gX$ and the eigenstates are product states polarized along the $x$-axis. The two eigendirections orthogonal to the Hamiltonian are $T_1\simeq Z$ and $T_2\simeq ZZ$. The distinction between them is particularly transparent in the $X$-basis. The operator $Z$ flips a single spin and therefore connects sectors with different total $x$-magnetization and a finite energy separation; however, a uniform $Z$ field simply rotates the independent spins and does not qualitatively alter the product-state structure of the eigenstates. Its associated susceptibility is therefore small. By contrast, $ZZ$ flips pairs of neighboring spins and connects states within the same highly degenerate $X$-magnetization sector. It therefore strongly hybridizes the degenerate manifold and qualitatively reorganizes the eigenstates, giving rise to a much larger susceptibility.

Combining these two limiting regimes allows us to interpolate the character of the two principal deformations $T_1$ and $T_2$ throughout the quantum-chaotic regime. In particular, we find that they are roughly represented by the orthonormal operators:
\begin{align}   
T_X &=\frac{J}{\varepsilon_*}\left[ \frac{gX+h Z}{\sqrt{g^2+h^2}}-\frac{\sqrt{g^2+h^2}}{J}ZZ\right], \nonumber \\
T_R &=\frac{hX-g Z}{\sqrt{g^2+h^2}}.
\label{eq:T1T2}
\end{align}
These expressions should not be regarded as the exact principal deformations of $G_{\alpha\beta}$, since finite-size effects, boundary conditions, and the choice of regularization can induce small rotations of the principal axes, causing $T_1$ and $T_2$ to have finite overlap with both $T_X$ and $T_R$. Nevertheless, they provide a useful physical interpretation of the two principal deformations, reduce to the asymptotic limits discussed above, and will also help clarify their behavior once inhomogeneous perturbations are introduced in Sec.~\ref{sec:hydro}. 

Specifically, $T_X$ is approximately aligned with the local magnetic field and primarily modifies the relation between the exchange term $J$ and the local magnetic field $\sqrt{g^2+h^2}$; $T_X \simeq ZZ $ when $g\rightarrow\infty$, while $T_X \simeq (JZ-hZZ)/\varepsilon_*$ when $g\rightarrow 0$. In contrast, $T_R$ is a field transverse to the magnetic field, and flips a spin in the direction of the field, thereby leading to a local jump in energy; $T_R \simeq Z$ when $g\rightarrow \infty$, whereas $T_R \simeq X$ when $g\rightarrow 0$. Thus, whereas $T_R$ predominantly induces large local energy-changing transitions, $T_X$ instead has a large overlap with the local energy density $h_i = gX_i+hZ_i+JZ_iZ_{i+1}$.  
One may therefore expect $T_R$ to produce a stronger rearrangement of the eigenbasis and hence a larger susceptibility. As such, in the most chaotic regime, the less susceptible principal perturbation $T_1$ is approximately aligned with $T_X$, while the slightly more susceptible one is more closely aligned with $T_R$, consistent with what we find by directly inspecting the eigenvectors of $G_{\alpha\beta}$ obtained numerically. 

The distinction between $T_X$ and $T_R$ will acquire a complementary dynamical interpretation in Sec.~\ref{sec:hydro}, when we consider spatial deformations of the operators. In particular, at finite momentum, $T_X$ can hybridize with the hydrodynamic energy mode because of its overlap with the local energy density, whereas $T_R$ remains much more weakly coupled to the hydrodynamic sector.

\begin{figure}
    \centering
    \includegraphics{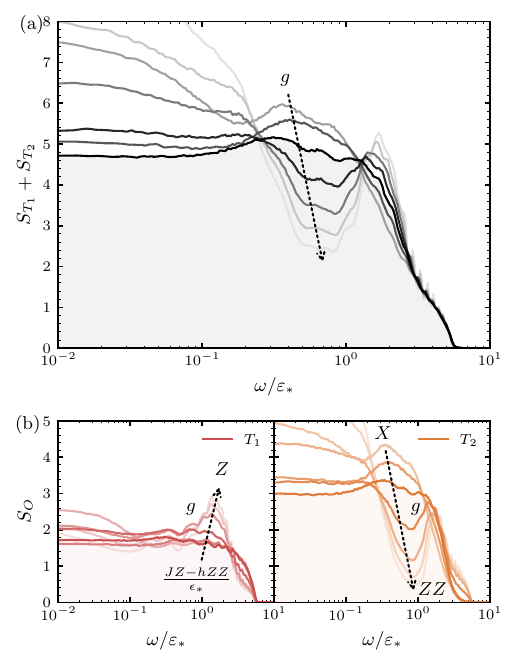}
    \caption{Spectral response along the principal deformations. (a) Total spectral function $S_{T_1}+S_{T_2}$ and (b) individual contributions $S_{T_1}$ (left) and $S_{T_2}$ (right), shown for different $g$ along the $h=0.3$ line cut. 
    The arrows indicate spectral responses for increasing $g$ along the line cut. The operators highlighted at both ends of the arrows indicate the asymptotic operators at the two limits $g\rightarrow 0$ and $g\rightarrow\infty$. Near the maximally chaotic regime, the spectral weight is broadly distributed over frequency, with strongly reduced low-frequency structure. Away from this regime, pronounced low- (and high-) frequency features develop, giving rise to an enhanced geometric susceptibility through the inverse frequency weighting in Eq.~(\ref{eq:geometrictensor}). Thus, the minimum of the total susceptibility corresponds dynamically to a broad dissipative response with minimal non-hydrodynamic low-frequency structure. The results are obtained for $J=1$, $N=14$, and by averaging over the central $10\%$ eigenstates.}
    \label{fig:gdependence}
\end{figure}

\subsection{Spectral Functions of the Eigendeformations}

Having described the principal deformations, we now use the connection between the geometric susceptibility and spectral functions in Eq.~(\ref{eq:geometrictensor}). In Fig.~\ref{fig:gdependence}(a), we show the total spectral function, obtained by summing the spectral functions associated with the two eigendeformations orthogonal to the Hamiltonian, along the same line cut in parameter space considered above. The lower subpanels resolve the contributions from each eigendirection separately. Curves near the susceptibility minimum at $g_*\approx1.05$ are highlighted in dark colors, with progressively lighter shades indicating increasing distance from $g_*$ on both sides.

As anticipated from the discussion in Sec.~\ref{sec:susceptibility}, the spectral functions associated with the principal deformations are both broadest and smoothest near the maximally chaotic point. In this regime, the low-frequency spectral weight is strongly suppressed, while the response is broadly distributed over frequencies of order of the microscopic energy scale, consistent with the small AGP susceptibility. The orthogonal operators $T_1$ and $T_2$ are normalized such that ${\rm Tr}[T_\alpha^\dagger T_\alpha]=N$, with only a negligible fraction of this weight residing in their diagonal matrix elements. Thus, the observed broadening reflects a redistribution of approximately fixed off-diagonal spectral weight over a wider frequency range, rather than a decrease in the total spectral weight as the Hamiltonian parameters vary.

As the system is tuned away from this point toward the integrable limits, additional structure develops in the spectral functions, most prominently in the form of sharp peaks associated with increasingly well-defined transitions between discrete energy levels. These features reflect the reemergence of integrable structure in both the Ising and paramagnetic limits.

The structure of the spectral functions near the integrable limits can be understood using the same microscopic picture discussed in the previous subsection. 
We first consider the principal deformations $T_1$ with the smaller susceptibility, whose limiting forms are $T_1\simeq (JZ-hZZ)/\varepsilon_*$ as $g\rightarrow0$ and $T_1\simeq Z$ as $g\rightarrow\infty$. Near the classical Ising limit, $H \simeq hZ+JZZ$, the perturbation $JZ-hZZ$ is diagonal in the classical eigenbasis and therefore carries no off-diagonal spectral weight at $g=0$. At small but finite $g$, it acquires only weak spectral weight through the perturbative dressing of the classical eigenstates. Correspondingly, the spectral response of $T_1$ remains weak, nearly flat, and largely featureless over the frequency range shown in the left panel of Fig.~\ref{fig:gdependence}(b). 
Near the opposite, paramagnetic limit, $H\simeq gX$, the operator $Z$ flips a single spin in the $X$ basis and therefore generates transitions at $\omega=\pm2g$, $S_Z(\omega)\propto\delta(\omega-2g)+\delta(\omega+2g)$, so that well-defined finite-frequency peaks begin to emerge as $g\rightarrow\infty$. The small spectral weight when $g\rightarrow 0$, and the finite frequency peaks when $g\rightarrow \infty$, make the low frequency weighting small in both limits.

We next turn to the principal deformation $T_2$ with larger susceptibility, for which $T_2\simeq X$ as $g\rightarrow0$ and $T_2\simeq ZZ$ as $g\rightarrow\infty$. Near the classical Ising limit, $H\simeq hZ+JZZ$, the operator $X$ flips a single spin. The corresponding energy change depends on its local environment through $h+J(Z_{i-1}+Z_{i+1})$, giving three characteristic transition energies, $\omega=2h$, $2(h+2J)$, $2(h-2J)$, together with their negative-frequency counterparts. Thus, in the strict classical limit, $S_X(\omega)$ consists of three pairs of sharp peaks whose relative weights are set by the probabilities of the corresponding local spin environments. At small but finite $g$, these discrete transitions broaden, as observed in Fig.~\ref{fig:gdependence}(b). Near the paramagnetic limit $H\simeq X$, by contrast, $T_2\simeq ZZ$ flips two neighboring spins in the $X$ basis. For oppositely aligned spins, this process leaves the total $X$ magnetization, and hence the energy, unchanged, whereas for aligned spins it changes the energy by $\pm 4 g$. Consequently, $S_{ZZ}(\omega)\propto \delta(\omega) +\frac{1}{2}\delta(\omega-4g) +\frac{1}{2}\delta(\omega+4g)$. These finite-frequency peaks are clearly visible for $g\gg g_*$ in the right panel of Fig.~\ref{fig:gdependence}(b). The enhanced low-frequency spectral weight of $T_2$ near both integrable limits yields a parametrically larger susceptibility than that of $T_1$, as shown in Fig.~\ref{fig:Geig}. Moving away from either limit broadens these sharply resolved microscopic transitions, eventually producing the broad, smooth spectral functions characteristic of the chaotic regime.

\section{Beyond the control Manifold: Hydrodynamics and Dissipation}
\label{sec:hydro}

So far, we have established the first link in Eq.~(\ref{eq:hypothesis}), connecting minimal geometric susceptibility with maximal eigenstate randomness, and have partially established the second link by showing that operators within the subspace orthogonal to the Hamiltonian exhibit broad, smooth spectral continua near the most chaotic regime. 

In this section, we strengthen the second link by extending the analysis beyond the operators contained in the control manifold $L_\alpha$. If the regime identified by minimal geometric susceptibility is genuinely maximally chaotic, all parametrically slow structure should be due to hydrodynamic modes, which are protected by locality, symmetries, and conservation laws, while generic operators orthogonal to this hydrodynamic sector should relax on microscopic timescales and thus exhibit broad spectral response.

To show that this is the case, we consider two classes of operators: 
(i) inhomogeneous deformations constructed from the same local building blocks $L_\alpha$ but with spatially varying coefficients, and (ii) operators that lie outside the original control manifold.

\subsection{Inhomogeneous Control Operators}

We first consider spatially inhomogeneous deformations generated from the same local operators $L_\alpha$ that span the control manifold. To this end, it is convenient to switch from open to periodic boundary conditions. Whereas the previous sections were designed to enable a direct comparison with earlier work~\cite{Rodriguez-Nieva_24}, periodic boundaries allow us to organize spatially inhomogeneous operators by momentum and thereby analyze the spectral response of the full set of $3L$ operators in momentum space by changing the wavevector $q$. 

We thus define the momentum-space components of the control operators as
\be
L_{\alpha,q}=\sum_{i=1}^{N} e^{-iqr_i}L_{\alpha,i}, 
\ee
where $L_{\alpha,i}$ denotes the operator $\alpha = X,Z,ZZ$ evaluated at lattice site $i$. Similarly, for the principal operators $T_\alpha$, we define $T_{\alpha,q}=\sum_i e^{-iqr_i}T_{\alpha,i}$, where the $q=0$ components are $T_1$ and $T_2$ considered in the previous section. 
In addition, we include the momentum-resolved Hamiltonian density, $H_q=\sum_i e^{-iqr_i}h_i$, which overlaps strongly with the hydrodynamic mode associated with energy conservation. 

\begin{figure}
    \centering
    \includegraphics{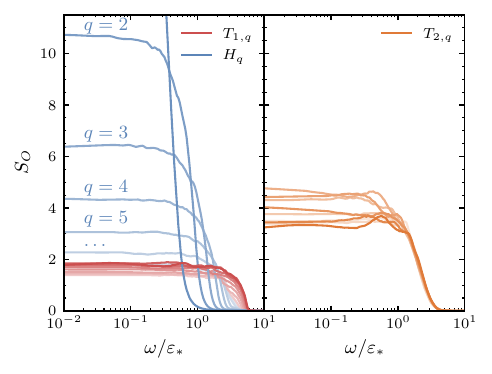}
    \caption{Momentum-resolved spectral response at $g=1.05$ and $h=0.3$. (Left panel) Spectral functions of $H_q$ and the least-susceptible principal perturbation $T_{1,q}$ for several momenta $q$, with lighter shades denoting larger $q$. At small $q$, the hydrodynamic mode is sharply concentrated at low frequencies, while the orthogonal mode remains broad. With increasing momentum, the two modes hybridize, and the distinction between hydrodynamic and non-hydrodynamic response becomes progressively less sharp. (Right panel) Spectral function of $T_{2,q}$, corresponding to the principal direction with largest susceptibility, which remains well separated from the hydrodynamic sector over the range of momenta shown and retains a broad, weakly $q$-dependent response. The results are for periodic boundary conditions with $J=1$, and $N=16$, and by averaging over the central $10\%$ eigenstates.}
    \label{fig:qdependence}
\end{figure}

The expectation is that, in strongly quantum-chaotic regimes, hydrodynamic relaxation provides the only parametrically slow relaxation mechanism. Generic operators will typically have a small or no overlap with hydrodynamic modes, while their spectral functions remain predominantly broad and governed by microscopic relaxation scales. By contrast, the operator $H_q$ has a strong overlap with the hydrodynamic energy mode and therefore develops a large low-frequency response at small $q$, diverging at small $q$, with a linewidth set by the hydrodynamic decay rate, $\Gamma_{q,{\rm hydro}}\simeq Dq^2$ [cf.~Eq.~(\ref{eq:scaling})].

In Fig.~\ref{fig:qdependence}, we show the corresponding spectral functions at the same parameter values $(g_*,h_*)$ identified above as maximally chaotic for all momenta $q=0,1,\ldots,N/2$ (while remaining wavevectors are related by inversion symmetry). In the left panel, we compare the hydrodynamic mode $H_q$ with the orthogonal deformation $T_{1,q}$. As anticipated, $H_q$ develops a pronounced low-frequency peak at small momentum reflecting the hydrodynamic pole $\Gamma_{\bm\theta,\rm hydro} \approx Dq^2$, while $T_{1,q}$ remains broadly distributed in frequency regardless of the value of $q$. Interestingly, the two responses $H_q$ and $T_{1,q}$ become strongly hybridized as $q$ approaches the lattice scale, $q\sim N/2$. Recalling that we can approximate $T_1 \approx T_X$ in Eq.~(\ref{eq:T1T2}), this reflects the fact that, at short wavelengths, the local densities $h_i$ and $T_{1,i}$ have large local overlap, so the distinction between hydrodynamic and orthogonal sectors becomes progressively less sharp at large $q$. 

Separately, in the right panel we show the spectral function of $T_{2,q}$, which changes only weakly with momentum and remains broad across the full range of $q$. Locally, this operator corresponds to a magnetic field transverse to the local fields entering the MFIM, and therefore has finite but small overlap with hydrodynamic modes.

Taken together, these results show that, among the $3N$ operators considered, pronounced low-frequency structure emerges only for operators overlapping with energy deformations, while the remaining operators retain broad spectral functions irrespective of $q$.

\begin{figure}
    \centering
    \includegraphics{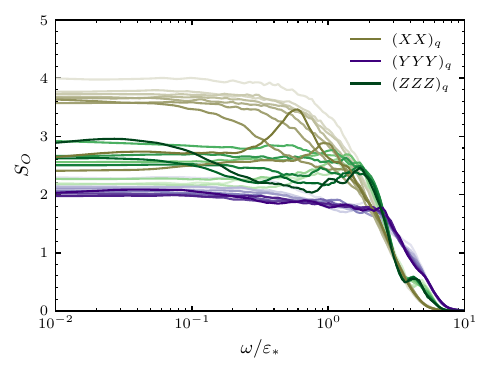}
    \caption{Spectral response of operators outside the Hamiltonian control manifold. We show the spectral function as a function of the normalized frequency $\omega/\varepsilon_*$ for $O=(XX)_q$, $O=(YYY)_q$, and $O=(ZZZ)_q$ for all possible $q$ values at $(g,h)=(1.05,0.3)$ for $N=16$ with periodic boundary conditions. $q=0$ is the darkest color for each operator with lighter shades indicating increasing values of $q\rightarrow \pi$. The spectral weight is broadly distributed over frequency with reduced low-frequency structure for all $q$ values. All results are obtained by averaging over the central $10\%$ eigenstates.}
    \label{fig:xx-yy}
\end{figure}

\subsection{Operators beyond the control manifold}

We now consider local operators that do not correspond to accessible Hamiltonian deformations, both uniform and inhomogeneous. These operators serve as independent probes of the dynamics, allowing us to ask whether the regime identified as maximally chaotic within the original three-dimensional control manifold also appears structureless to more generic local probes. We emphasize that these operators are not added to the control manifold itself, in which case the susceptibility geometry and its minimum would generally change.

Generic local operators outside the manifold of allowed Hamiltonian deformations can---and typically do---overlap with hydrodynamic modes. Indeed, even when their linear overlap with the energy density vanishes (as occurs for operators outside the allowed Hamiltonian manifold), nonlinear couplings to conserved densities can still generate hydrodynamic long-time tails~\cite{PhysRevX.15.011059}. Nevertheless, these contributions do not necessarily carry a substantial fraction of the total spectral weight, as much of that weight typically resides in a nonhydrodynamic component that relaxes on microscopic timescales, producing a broad continuum that coexists with the slow hydrodynamic response.
The relative weight of these components varies strongly between operators. For the MFIM, for example, \(Y_q\propto[H,Z_q]\), and therefore \(S_Y(q,\omega)\propto\omega^2S_Z(q,\omega)\), producing a vanishingly small spectral weight near \(\omega=0\). Related low-order commutator descendants exhibit analogous structured spectral functions.

The nearest-neighbor \(XX\) operator provides an example with much weaker overlap with the hydrodynamic sector and appears only at relatively high order (order four) in the local commutator hierarchy. Its spectral function is correspondingly broad and exhibits only weak momentum dependence over the accessible system sizes, as shown in Fig.~\ref{fig:xx-yy}. Similar behavior is observed for generic three-site operators. In particular, both $YYY$ and $ZZZ$ predominantly exhibit broad spectral continua, although their detailed spectral shapes differ because they belong to different symmetry sectors, with $YYY$ odd and $ZZZ$ even under complex conjugation.

These results support a stronger interpretation of the susceptibility minimum found within the control manifold. Despite the MFIM being specified by only three tunable couplings, its chaotic regime appears structureless to a much larger class of local probes: pronounced low-frequency features are observed only from strong overlap with conserved, hydrodynamic modes, whereas operators largely orthogonal to these modes exhibit broad microscopic relaxation. This supports our original expectation that 
the only remaining slow structure is that imposed by hydrodynamics.

\section{Discussion}
\label{Sec:Discussion}

Our results reveal that eigenstate susceptibility, eigenstate randomness, broad and smooth spectral response, and asymptotically slow hydrodynamics are intimately connected through a common underlying geometric structure, thus providing a unified description of distinct facets of quantum chaos. From this perspective, the strong mixing of eigenstates under physical deformations, the broadening of the spectral functions of local operators, and the suppression of overlap with low-frequency structures are different manifestations of the same underlying geometry, as encoded in Eq.~(\ref{eq:geometrictensor}). Each probes, in a complementary way, how much physical structure is retained in the eigenstates---or, conversely, how thoroughly such structure has been erased. 

One consequence of this connection is that our work provides a simple geometric interpretation of regimes previously identified as {\it maximally chaotic}~\cite{Rodriguez-Nieva_24}. In these regimes, all structure not protected by physical constraints, including locality, symmetries, and conservation laws, is effectively washed out, such that no physically allowed Hamiltonian deformation can further remove any residual eigenstate structure.
From a dynamical perspective, this leaves hydrodynamics as the only source of parametrically slow relaxation. Our framework thus provides a quantitative criterion for maximizing chaos that can be applied directly to generic quantum many-body systems, without relying on semiclassical limits~\cite{Kitaev_18,Polchinski_16,Roberts_18}.

There are several interesting ramifications of our work. First, the present analysis brings together a broad and important set of signatures of quantum chaos, including eigenstate sensitivity, randomness, spectral broadening, and hydrodynamic structure within a single framework. These manifestations are complemented by our previous results showing that the same regimes are singled out by how fast randomness emerges under unitary Hamiltonian dynamics~\cite{ghosh_25}. The signatures unified here, however, represent a small part of a much broader web of quantum-chaos diagnostics, including scrambling, pseudorandomness generation, and a growing set of fine-grained statistical probes~\cite{Pandey_20,2022Quantum_designs,2022PRL_exactdesigns,Rodriguez-Nieva_24,gjonbalaj_choi_26,Pappalardi_25}. 
An important question is whether the geometric framework developed here can unify this much broader web of chaos diagnostics. 
Our results suggest that it can, as our formalism directly quantifies the residual physical structure of many-body eigenstates, which in turn underlies and constrains all these seemingly distinct dynamical manifestations of quantum chaos. 

A second intriguing implication is that quantifying chaos opens the possibility of establishing bounds on how chaotic a many-body system can become~\cite{PhysRevLett.123.230606,2019PRX_lanczos,Maldacena_16,PhysRevX.9.011034}, and studying how these bounds change under different physical constraints. Interestingly, such bounds would simultaneously constrain the amount of eigenstate randomness, eigenstate mixing, and dynamical relaxation within a single framework; it could also help clarify the conditions under which extremal behavior can be realized, and how generic these behaviors are across many-body systems. 

While the present work focuses on quantum spin chains and a particular manifold of local Hamiltonian terms, our broader conjecture is that the same geometric picture extends to generic local many-body systems. Natural questions include how this picture is modified by additional symmetries and conservation laws~\cite{2023PRL_su2chaitanya}, kinematic constraints~\cite{PhysRevX.12.011050}, anomalous hydrodynamics~\cite{PhysRevX.11.031023}, or long range interactions~\cite{PhysRevB.101.020416}. These extensions could help establish the generality of the geometric framework and clarify the extent to which different manifestations of maximal chaos share a common physical origin.

Beyond fundamental questions of quantum chaos, our framework has natural applications to programmable quantum platforms with restricted control capabilities. In such systems, many-body quantum dynamics is inherently tied to the experimentally accessible controls. The geometric tensor provides a quantitative characterization of this manifold, identifying directions along which the many-body eigenbasis is most sensitive to changes in control parameters, as well as directions along which it remains comparatively robust. This geometric perspective could guide the identification of optimal operating regimes for efficient quantum state randomization~\cite{Brandao_16,foxman_25}, robust quantum control~\cite{PhysRevLett.132.193801}, counterdiabatic driving~\cite{Morawetz_25}, and scrambling-assisted many-body sensing~\cite{kobrin_24,Hu_26}. 

\section*{Acknowledgements}

We thank Anatoli Polkovnikov for stimulating discussions. The authors acknowledge the hospitality of the Kavli Institute for Theoretical Physics through the program {\it Learning the Fine Structure of Quantum Dynamics in Programmable Quantum Matter}, supported by NSF Grant No.\ PHY-2309135. JRN also gratefully acknowledges support from the Simons Center for Geometry and Physics, Stony Brook University at which some of the research for this paper was performed. Numerical simulations were performed using the advanced computing resources provided by Texas A\&M High Performance Research Computing.

\appendix
\renewcommand{\thefigure}{A\arabic{figure}}
\setcounter{equation}{0}
\setcounter{figure}{0}

\section{Control Operators for Open Boundary Conditions}
\label{app:OBC}

We give here the explicit control operators used for open boundary conditions, where we add fields $h_bZ_1$ and $-h_bZ_N$ at the boundaries to break inversion symmetry. We choose their normalization such that $(1/\mathcal D)\operatorname{Tr}(L_\alpha^\dagger L_\beta)=N\delta_{\alpha\beta}$. The three operators are
\be
\begin{aligned}
L_X &= \sum_{i=1}^{N} X_i,\\
L_Z &= \frac{1}{\sqrt{1+2h_b^2/(h^2N)}}
\left[
\sum_{i=1}^{N} Z_i+\frac{h_b}{h}(Z_1-Z_N)
\right],\\
L_{ZZ} &= \frac{1}{\sqrt{1-1/N}}
\sum_{i=1}^{N-1} Z_iZ_{i+1}.
\end{aligned}
\ee
Here, the missing boundary bond modifies the normalization of $L_{ZZ}$, while the inversion-breaking boundary fields $h_b$ are absorbed into $L_Z$. 
With these definitions, one can map the Hamiltonian $H_{\bm\theta}$ in Eq.~(\ref{eq:Ham}) to the MFIM in Eq.~(\ref{eq:MFIM}) via 
\be
\begin{aligned}
   \theta_X &= g, \\ 
   \theta_Z &= h\sqrt{1+{2h_b^2}/{h^2N}}, \\ 
   \theta_{ZZ} &= J\sqrt{1-1/N},
\end{aligned}
\ee
with energy $\varepsilon_* = \sum_\alpha\theta_\alpha^2$ given by $\varepsilon_* = \sqrt{J^2+g^2+h^2+\frac{2h_b^2-J^2}{N}}$.

\section{Dependence of the Geometric Susceptibility on the Regularization Scale}
\label{app:regularization}

\begin{figure}
    \centering
    \includegraphics{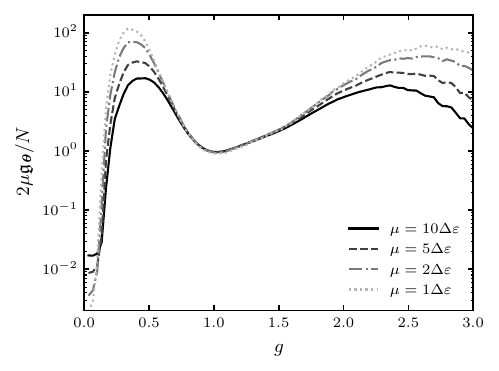}
    \vspace{-5mm}
    \caption{Regularization dependence of the total geometric susceptibility. We rescale the susceptibility as $2\mu{\mathfrak g}_{\bm\theta}/N$, along the $h=0.3$ line cut, for different $\mu$ to make it independent of the regularization scale in the chaotic regime 
    and, according to Eq.~(\ref{eq:scaling}), it corresponds to the inverse characteristic spectral broadening $\Gamma_{\bm\theta}^{-1}$. 
    The excellent collapse of the curves in the chaotic regime shows the independence of the total geometric susceptibility on the choice of the regularization scale. The results are for a chain of $N=14$ sites with $J=1$ and are averaged over the central $10\%$ eigenstates.}
    \label{fig:mu_dependence}
\end{figure}

The results reported in the main manuscript use the regularization $\mu=5\,\Delta\varepsilon$ both for the AGP and the Lorentzian approximation of the $\delta-$functions in the definition of the spectral function. Here, we show that, as long as $\mu$ remains within the limits explained in the main manuscript, the choice of $\mu$ does not have substantial impact on the qualitative behavior of the AGP. Furthermore, the choice of $\mu$ does not quantitatively change the physics observed within the chaotic regime, our regime of interest here. In Fig.~\ref{fig:mu_dependence}, we plot the rescaled total geometric susceptibilities $2\mu{\mathfrak g}_{\bm\theta}/N$ for different values of the regularization $\mu=a\,\Delta\varepsilon$, where $a\sim \mathcal{O}(1)$ to remain within the same regime identified in the main manuscript $\mu \ll D/N^2$. The rescaled susceptibilities show perfect collapse in the regime $0.5 \lesssim g \lesssim 2$, indicating that the choice of $a$ does not affect the calculated $\Gamma_{\bm\theta}$ in this regime, where quantum chaos is established in this finite system with $N=14$. The departure of curves from this collapse at different $a$ values reflects that the model is not ergodic at this finite system size, where the dependence of the susceptibility on $\mu$ is different. Lowering $\mu$ gives rise to faster divergences near the integrable regimes since it becomes less efficient in suppressing spectrum (near) degeneracies typical for integrable models. However, the qualitative features of the total susceptibilities do not change for the models and system sizes considered here.

We also have explicitly checked that the choice of $\mu$ used to regularize the $\delta$-function appearing in the spectral functions has no appreciable quantitative effect on the spectral response. Varying $a$ in $\mu=a\,\Delta\varepsilon$ produces imperceptible changes in the plots already shown in Fig.~\ref{fig:gdependence}. Thus, provided $\mu\ll D/N^2$, the spectral functions are largely insensitive to the precise choice of regularization.

\begin{figure}
    \centering
    \includegraphics{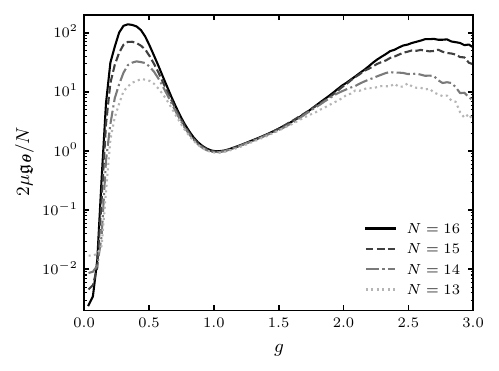}
    \vspace{-5mm}
    \caption{Size-dependence of the total geometric susceptibility. We rescale the susceptibility as $2\mu{\mathfrak g}_{\bm\theta}/N$ to make it independent of the regularization scale in the chaotic regime and, according to Eq.~(\ref{eq:scaling}), it corresponds to the inverse characteristic spectral broadening $\Gamma_{\bm\theta}^{-1}$, along the $h=0.3$ line cut, vs $g$ for different chain lengths $N$. The excellent collapse of the curves in the chaotic regime shows the independence of the total geometric susceptibility on the system size. The results are for chains with $\mu=5\,\Delta\varepsilon$, $J=1$ and are averaged over the central $10\%$ eigenstates.}
    \label{fig:N_dependence}
\end{figure}

\section{Finite-size scaling of the \\ Geometric Susceptibility}
\label{app:finitesizescaling}

Next, we show the dependence of the total geometric susceptibilities on the system size. We plot the rescaled susceptibilities for different system sizes across the $h=0.3$ line cut in Fig.~\ref{fig:N_dependence}. Again, we observe excellent collapse of the curves from different system sizes in the chaotic regime, showing the robustness of the results reported in the main manuscript. The extracted spectral broadening $\Gamma_{\bm\theta}$ in the chaotic regime is robust to changes in system sizes, and the regime where it can be robustly extracted gets extended with increasing system size. Additionally, the location of the susceptibility minimum $(g_*,h_*) \approx (1.05, 0.3)$ does not seem to move appreciably with the system size, while the minimum gets sharper with increasing system size. The peaks to the right and left of the maximally chaotic point indicate the crossover between ergodicity and integrability for the corresponding system size. They shift towards the integrable limits $g=0$ and $g\rightarrow\infty$ with increasing system size, as expected, making evident that the dark regions close to the boundaries observed in Figs.~\ref{fig:AGPColormap},~\ref{fig:G_distribution} in the main manuscript are just finite-size effects that would be eliminated in the thermodynamic limit.

\bibliography{References}

\end{document}